\documentclass[twocolumn,showpacs,preprintnumbers,amsmath,amssymb,epsf,superscriptaddress]{revtex4-1}

\usepackage{graphicx,epsfig}
\usepackage{dcolumn}
\usepackage{bm}
\usepackage{color}

\begin{document}


\title{Transition between Macroscopic Steady Slippage and Creep Motion in a System with Velocity-Dependent Friction Stress}
\author{Takehito Suzuki}
\email{t-suzuki@phys.aoyama.ac.jp}
\author{Hiroshi Matsukawa}
\affiliation{Department of Physics and Mathematics, Aoyama Gakuin University, 5-10-1 Fuchinobe, Chuo-ku, Sagamihara, Kanagawa 252-5258, Japan}


\begin{abstract}
We investigate the propagation of a slip front in a visco-elastic body on a rigid substrate. The body is one-dimensional, and the loading stress is applied at one end. 
By employing a local friction law that has a quadratic form of the slip velocity and gives vanishing friction stress at vanishing velocity or above a certain velocity, we show analytically that macroscopic steady slippage and creep motion can be understood in a single framework. The critical values of the end-loading stress causing macroscopic steady slippage and the slip-front propagation velocity appear are obtained. 
These values are completely determined by the gradient of the slip velocity$-$friction curve at the vanishing friction stress. These results are extended to more general friction laws, and found to be consistent with numerical calculations. Furthermore, we discuss some seismological implications based on the analytical and numerical results.
\end{abstract}

\maketitle

\section{Introduction} \label{secI}

When we apply a small loading stress in a tangential direction to a solid object on a solid substrate and increase the force slowly, the block slips as a whole only when the force exceeds a critical value, i.e., the macroscopic static friction force. Though this behavior has long been recognized, the condition determining whether or not macroscopic steady slippage occurs has not been clarified in a single framework. Moreover, local slippages, called precursors, are observed in some systems prior to macroscopic sliding (e.g., \cite{Ben, Mae, Ots, Lan93, Rad, Tro, Amu15, Amu12, Bar}). These precursors show slip velocity and stress concentration negligibly small to those of the macroscopic steady slip, and can be arrested spontaneously or by small local perturbations, such as a small increase of the normal stress. From physical, industrial, and seismological viewpoints, it is necessary to understand the relationship between the precursors and the initiation of the macroscopic steady slippage to clarify the mechanism for macroscopic slip initiation.

Geophysical studies have also contributed to our understanding of the  behavior of macroscopic slip initiation. 
In particular, slow earthquakes, which have slip velocities and fault-tip-propagation velocities that are negligibly small compared to those of ordinary earthquakes, have been observed \cite{Oba}. The relationships among slow earthquakes and ordinary earthquakes$-$in particular, whether or not slow earthquakes can trigger ordinary earthquakes$-$is currently a controversial problem.

Macroscopic slip initiation has also been investigated experimentally using, e.g., soft gels \cite{Rub, Bau02, Bau03, Yam09}. In particular, viscosity is considered to be important for such gel experiments to produce earthquake-like slippage. However, from an analytical viewpoint, the effect of the viscosity on the macroscopic slip initiation$-$in particular, on the slip-front propagation velocity$-$remains unclear. 

Here, we consider slip-front propagation (SFP) into an intact, homogeneous area under the condition of constant loading stress at one end of the system to understand macroscopic slip initiation. This end-loading model has been used to treat the behavior of macroscopic slip initiation (e.g., \cite{Ben, Mae, Ots, Rad, Amu15, Tro}). If steady SFP with a finite propagation velocity is observed, we consider that macroscopic steady slippage occurs. If such steady SFP is not observed, we regard the slip as creep motion, because the slip velocity decays with increasing time and macroscopic steady motion does not occur. We investigate the slippage of a visco-elastic block on a rigid substrate

We employ a local friction law that depends quadratically on the slip velocity and gives vanishing friction stress above a certain velocity, and obtain the exact value for the minimum velocity of SFP.
We also find a critical value of the applied stress, above which the SFP has a finite propagation velocity; below the critical value, the SFP decays. This means that the transition between macroscopic steady slippage and creep motion appears in the model with the local friction law employed here.
In addition, the physical consideration and the Linear Marginal Stability Hypothesis show that the analytical results are not specific to the friction law noted above but are applicable to more general friction laws as well. The gradient of the slip velocity$-$friction stress curve at the vanishing friction stress plays an important role for such discussion. Finally, we investigate SFP in visco-elastic systems numerically and have obtained results that are consistent with the analytic ones.
In the numerical calculations, we extend the present framework to the friction law with always positive friction stress. Based on these results, we can understand, in terms of the local friction law, whether or not macroscopic steady slippage occurs and how front propagation is determined. Implications of this work for precursors and slow earthquakes based on slippage with a driving stress less than the critical value are presented.

\section{Visco-Elastic Model with a Velocity-Dependent Local Friction Law} \label{secMWVD}

\subsection{Model and Definition of Slip-Front Propagation} \label{secMod}

We consider a visco-elastic block on a rigid and fixed substrate and apply the loading stress acting on the left end in the direction tangential to the substrate surface, i.e., end-loading stress. The block is assumed to be an infinite homogeneous medium that is a one-dimensional (1D) system in the $x$ direction. This condition is realized as follows. Let us consider a 3D block, and push the end-loading point whose height from the substrate is $h$. We can regard the system as 1D in the region where the distance in the $x$ direction from the end-loading point is significantly larger than $h$, because the block can be regarded as an infinitely thin medium there. Deformation of the block is restricted to the $x$ direction.

At $t<t_0$, the block is homogeneous, and the slip distance $u(x,t)$ at position $x$ and time $t$ is equal to zero throughout the whole system. The constant end-loading stress is applied in the $x$ direction at time $t \ge t_0$, by pushing the block at $x=-\infty$, which yields a constant negative strain at that point. Here, we consider slip-front propagation (SFP) into an intact area, where $u(x,t)=0$, before the slip front arrives. Thus, we require the boundary conditions are given by
\begin{equation}
\lim_{x \to -\infty} \frac{\partial u}{\partial x} =p_{-\infty}(<0), \label{eqb1}
\end{equation}
and
\begin{equation}
\lim_{x \to \infty} \frac{\partial u}{\partial x} = 0,  \label{eqb2}
\end{equation}
where $p_{-\infty}$ is a constant. 

The equation of motion is given by
\begin{equation}
\rho \frac{\partial^2 u}{\partial t^2} =E_1 \frac{\partial^2 u}{\partial x^2}+\eta \frac{\partial^3 u}{\partial x^2 \partial t}-\tau_{\mathrm{fric}}, \label{eqeom0}
\end{equation}
where $\rho$ is the mass density, $E_1$ is Young's modulus, $\eta$ is the viscosity of the block, and $\tau_{\mathrm{fric}}$ is the local friction stress. 
If we use the characteristic length $L_0$ and time $T_0$, Eq. (\ref{eqeom0}) becomes
\begin{equation}
\frac{\partial^2 \tilde{u}}{\partial \tilde{t}^2} =\frac{E_1 T_0^2}{\rho L_0^2} \frac{\partial^2 \tilde{u}}{\partial \tilde{x}^2}+\frac{\eta T_0}{\rho L_0^2} \frac{\partial^3 \tilde{u}}{\partial \tilde{x}^2 \partial \tilde{t}}-\frac{T_0^2}{\rho L_0} \tau_{\mathrm{fric}},
\end{equation}
where $\tilde{u}, \tilde{x}$ and $\tilde{t}$ are the nondimensionalized slip displacement, space, and time, respectively. 
We put $E_1 T_0^2/\rho L_0^2 \equiv 1$ and $\eta T_0/\rho L_0^2 \equiv 1$, which gives elastic wave with velocity unity. From Eqs. (\ref{eqb1}) and (\ref{eqb2}), the boundary conditions then become
\begin{equation}
\lim_{\tilde{x} \to -\infty} \frac{\partial \tilde{u}}{\partial \tilde{x}} =\tilde{p}_{-\infty} =p_{-\infty},
\end{equation}
and
\begin{equation}
\lim_{\tilde{x} \to \infty} \frac{\partial \tilde{u}}{\partial \tilde{x}} = 0, 
\end{equation}
where $\tilde{p}_{-\infty}$ is the normalized strain, which equals $p_{-\infty}$. Henceforth, we describe the nondimensionalized values $\tilde{u}$, $\tilde{x}$ and $\tilde{t}$ as $u$, $x$ and $t$, respectively. The equation of motion thus becomes
\begin{equation}
\ddot{u} = u'' + \dot{u}'' - \tau, \label{eqeom1}
\end{equation}
where the dot and prime represent differentiation with respect to time and space, respectively, and $\tau \equiv T_0^2 \tau_{\mathrm{fric}}/\rho L_0$ is the normalized local-friction stress. Note that the constant $p_{-\infty}$ represents both the constant stress at the position $x \to -\infty$, which equals to the strain at that position. We use the word ``stress'' instead of ``strain'' henceforth. 

\begin{figure}[tbp]
\centering
\includegraphics[width=7.5cm]{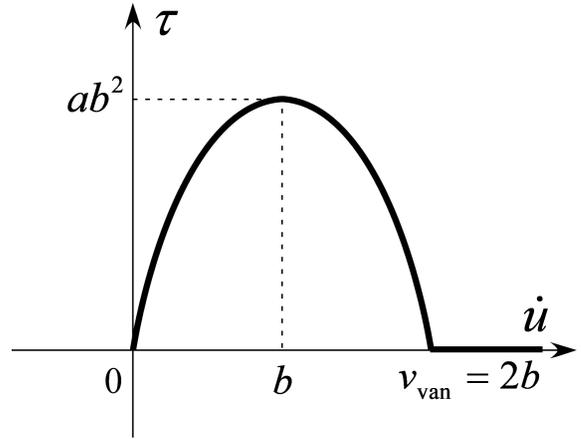}
\caption{Constitutive law of the local friction stress as a function of  the slip velocity.}
\label{FigCL}
\end{figure}

At first, we assume that $\tau$ takes the form
\begin{eqnarray}
\tau &=& a \dot{u} (2b -\dot{u}) [H(\dot{u})-H(\dot{u}-2b)] \nonumber \\
&=& -a \dot{u} (\dot{u}-v_{\mathrm{van}}) [H(\dot{u})-H(\dot{u}-v_{\mathrm{van}})],  \label{eqtau}
\end{eqnarray}
in order to obtain exact results analytically. Here $a$ and $b$ are positive constants, $v_{\mathrm{van}}=2b$ is the critical value of the slip velocity above which $\tau$ vanishes, $H(\cdot)$ is the Heaviside function (see also Fig. $\ref{FigCL}$). With this local friction law, the friction stress changes from a velocity-strengthening to a velocity-weakening behavior at $\dot{u}=b$ with increasing slip velocity, and it vanishes for $\dot{u} \ge 2b =v_{\mathrm{van}}$. 
%

We define the position of the slip front as the position where $\tau$ reaches a given small reference value after reaching its peak value $ab^2$, i.e., $\dot{u} \le v_{\mathrm{van}}$ and $\dot{u} \sim v_{\mathrm{van}}$, based on analogy with dynamic crack-tip propagation. The reference value does not affect the following study if it is negligibly small compared to $ab^2$. If the slip velocity decays with time over the whole system, we consider that SFP vanishes. When a slip front passes, $\tau$ and $\dot{u}$ at a fixed point $x$ change as follows. Before the slip front approaches, $\tau$ is zero because $\dot{u}(x,0)=0$. As the slip front approaches, $\dot{u}$ becomes positive, and $\tau$ increases with increasing slip. As $\dot{u}$ continues to increase, $\tau$ increases to its peak value $ab^2$ (when $\dot{u}=b=v_{\mathrm{van}}/2$) and then decreases to zero when $\dot{u}$ increases beyond $v_{\mathrm{van}}$. 

The model based on Eqs. (\ref{eqeom1}) and (\ref{eqtau}) is, in some sense, idealized. It is important, however, that an exact treatment can be performed with this model, as described in Sec. \ref{secExact2}. In fact, we obtain the exact values of the critical magnitude of the end-loading stress causing the steady SFP and of the SFP velocity. We will give two extensions of the friction law (\ref{eqtau}): (i) the friction law will be extended to laws with non-quadratic dependence on the slip velocity but friction stress vanishes above a certain critical velocity (see Secs. \ref{secExact2} and \ref{secLMSH}), and (ii) we will show that the analytical result obtained for the law is approximately valid even in the friction law where the friction stress approaches zero with $\dot{u} \to \infty$ (see Secs. \ref{secExact2} and \ref{secNum2}). 

We can give some bases qualitatively supporting Eqs. (\ref{eqeom1}) and (\ref{eqtau}). 
This friction law is characterized by the peak value $ab^2$ at $\dot{u}=v_{\mathrm{van}}/2$, and by zero friction stress for $\dot{u}=0$ and $\dot{u} \ge v_{\mathrm{van}}$. The velocity dependence of friction force which shows maximum is observed in laboratory experiments on soft matter \cite{Yam19}. The rock experiments show that the coefficient of kinetic friction exhibits a maximum for some systems and then almost vanishes above a certain slip velocity \cite{DiT}. In some seismological systems, thermal pressurization \cite{Suz06} or melting at the sliding plane \cite{Hir} also makes the sliding friction stress almost vanish in the large-velocity regime. The static friction stress can vanish due to thermal activation, and the friction stress can be expanded in terms of the power of the velocity in the small-slip-velocity regime \cite{Mat}. These statements support the local friction law employed here. We have also introduced the viscosity term, $\dot{u}''$, involving the time derivative of $u''$. This form is known as the Kelvin viscosity, and it has been employed in various works, e.g.,  \cite{Lan93, Mye1, Lan92, Sha, Mye2}.

\subsection{Analytical Study} \label{secAna}

\subsubsection{Exact Results for the SFP Velocity} \label{secExact2}

Here, we derive the front velocity for steady SFP analytically for the present visco-elastic system with the local friction law (\ref{eqtau}). The governing equation is given by
\begin{equation}
\ddot{u}= u'' + \dot{u}'' + a \dot{u} ( \dot{u} - v_{\mathrm{van}} ) [H(\dot{u})-H(\dot{u}-v_{\mathrm{van}})]. \label{eqeom2}
\end{equation}
We assume the solution to depend only on $x_1 \equiv x-vt$ with $v >0$. Note that $v$ is not a model parameter, but the value to be determined. With this assumption, Eq. (\ref{eqeom2}) leads to
\begin{equation}
(1 - v^2) u'' - v u''' - avu'(vu' +v_{\mathrm{van}}) [H(u')-H(u'+\frac{v_{\mathrm{van}}}{v})]=0, \label{eqSS1}
\end{equation}
where the prime here denotes the differentiation with respect to $x_1$. Here we define $q(X) \equiv (v/v_{\mathrm{van}}) (u'+v_{\mathrm{van}}/v)=-(\dot{u}-v_{\mathrm{van}})/v_{\mathrm{van}}$ and $X \equiv -x_1 \sqrt{a v_{\mathrm{van}}}$. Equation (\ref{eqSS1}) is expressed as
\begin{equation}
\frac{\partial^2 q}{\partial X^2} = -\mu \frac{\partial q}{\partial X} -\frac{\partial}{\partial q} \left( \frac{q^2}{2} - \frac{q^3}{3} \right) [H(q)-H(q-1)], \label{eqq2}
\end{equation}
where 
\begin{equation}
\mu=\frac{1-v^2}{\sqrt{a v_{\mathrm{van}}} v }. \label{eqmuv}
\end{equation}

\begin{figure}[tbp]
\centering
\includegraphics[width=8.5cm]{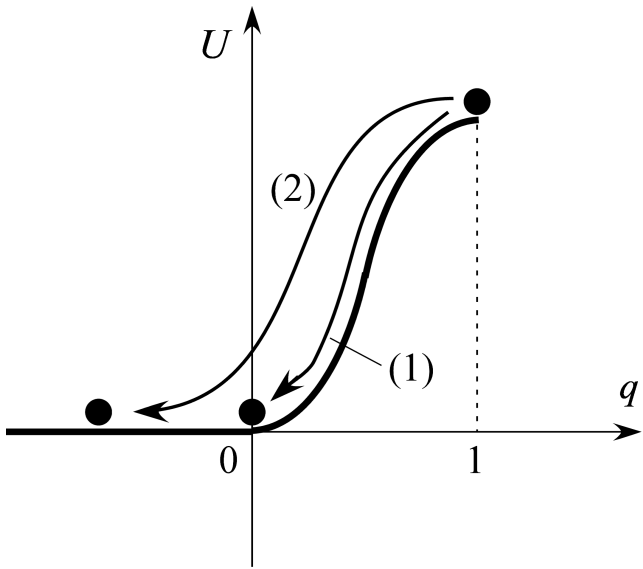}
\caption{Schematic illustration of the potential $U$. The black particle of unit mass obeys the equation of motion (\ref{eqq2}). Case (1) corresponds to $q(X \to \infty)=0$, while case (2) stands for $q(X \to \infty)<0$. The conditions $q=1$ and $q=0$ correspond to $\dot{u}=0$ and $\dot{u}=v_{\mathrm{van}}$, respectively.}
\label{FigSq}
\end{figure}

In Eq. (\ref{eqq2}) we can regard $q$ and $X$ as the displacement and time, respectively. Then Eq. (\ref{eqq2})  is just the equation of motion for a particle with unit mass under the potential $U=(q^2/2 - q^3/3)[H(q)-H(q-1)]$, with a damping force proportional to the velocity, $-\mu \partial q/\partial X$ (Fig. \ref{FigSq}). Here, $\mu$ corresponds to the damping constant. It is clear that $U$ is constant for $q \le 0$ and has an unstable point at $q=1$. First, we consider a solution starting from the unstable state $q(X \to -\infty)=1$, which corresponds to $\dot{u}=0$, to the right end point of the constant region $q(X \to \infty)=0$, which corresponds to $\dot{u}=v_{\mathrm{van}}$. The damping constant $\mu$ has a critical value corresponding to critical damping, above which the solution approaches $q=0$ monotonically [see case (1) in Fig. \ref{FigSq}]. In Eq. (\ref{eqq2}), the critical value, $\mu_c$, is given by $\mu_c=2$, as is easily shown. For the case of critical damping, we obtain $(1 - v^2)/(\sqrt{a v_{\mathrm{van}}} v) = 2$, which yields \cite{Aro}
\begin{equation}
v=\sqrt{1 + a v_{\mathrm{van}}} -\sqrt{a v_{\mathrm{van}}} \equiv v_c^{(-)}. \label{eqv-e}
\end{equation}
Here we have chosen only the positive-velocity solution. Note here that $a v_{\mathrm{van}}=\Big| \partial \tau/\partial \dot{u}|_{\dot{u}=v_{\mathrm{van}}} \Big| \equiv g$, and $g$ determines the propagation velocity as 
\begin{equation}
v_c^{(-)} =\sqrt{1+g}-\sqrt{g}. \label{eqv-eg}
\end{equation}
This will be confirmed from another viewpoint in Sec. \ref{secLMSH} and Appendix \ref{secAA}. 

For the boundary values $q(X \to -\infty)=1$ and $q(X \to \infty)=0$, we have $u'(x_1=x-vt \to \infty)=0$ and $u'(x_1=x-vt \to -\infty)=-v_{\mathrm{van}}/v_c^{(-)}$, respectively. These give the boundary values for a stress field, $p(x, t) \equiv \partial u/\partial x$; i.e., $p(x \to -\infty, t)=p_{-\infty}=-v_{\mathrm{van}}/v_c^{(-)}$ and $p(x \to \infty, t)=0$. 
From the relationship $\dot{u}=\partial u/\partial t =-v \partial u/\partial x_1=-vp$, we have $\dot{u}(x \to -\infty, t)=v_{\mathrm{van}}$ and $\dot{u}(x \to \infty, t)=0$. It is to be noted that $u$ is a function of $x_1=x-vt$. It turns out that a stable slip region with slip velocity $v_{\mathrm{van}}$ extends to the intact area when we apply the normal stress $p_c \equiv -v_{\mathrm{van}}/v_c^{(-)}$ at $x \to -\infty$. This corresponds to steady SFP moving to the right direction with the constant propagation velocity $v_c^{(-)}$.

Next, we consider the case $q(X \to \infty)<0$. See case (2) in Fig. \ref{FigSq}. The analytical treatment employed here can also be applied for this case. The work done by the damping force should be equal to the potential energy that the particle initially has \cite{Aro}. Therefore, the equation
\begin{equation}
\int_1^{q(X \to \infty)} \mu \frac{\partial q(X)}{\partial X} dq =U(1)=\frac{1}{6} \label{eqEconsv}
\end{equation}
must be satisfied. 
Here $q(X)$ is a function of $\mu$. Equation (\ref{eqEconsv}) is an equation for $\mu$, while it may not be analytically solved.

From Eq. (\ref{eqmuv}) and $v>0$, we have the relationship
\begin{equation}
v=\frac{\sqrt{a v_{\mathrm{van}} \mu^2 +4}-\mu \sqrt{a v_{\mathrm{van}}}}{2}. \label{eqSolv}
\end{equation}
Because the distance of motion of the particle is larger in the case (2) than in the case (1), $\mu<\mu_c$ must be satisfied for the case. Since $v$ is a monotonically decreasing function of $\mu$, $v>v_c^{(-)}$ in the case (2). 

The condition $q(X \to \infty) <0$ corresponds to $\dot{u}(x \to -\infty, t)>v_{\mathrm{van}}$. This means that the region where slip front has passed is moving with $\dot{u}>v_{\mathrm{van}}$; i.e., the region has overpassed the ``barrier'' of the friction stress in Fig. \ref{FigCL} and reached the state with vanishing friction stress $\tau$. In addition, note that $\dot{u}(x \to -\infty, t)$ increases monotonically with increasing $|p_{-\infty}|$. Therefore, we can conclude that if the magnitude of $|p_{-\infty}|$ is larger than $|p_c|$, which corresponds to $\dot{u}(x \to -\infty, t) =v_{\mathrm{van}}$, the barrier can be overpassed. 

If $q(X \to \infty) >0$, there is no stable solution because the particle in Fig. \ref{FigSq} cannot stop in the region $0<q<1$. The macroscopic steady slippage does not occur here, and the slip velocity becomes negligibly small far from the end-loading point. This case corresponds to $\dot{u}(x \to -\infty, t) <v_{\mathrm{van}}$ and $|p_{-\infty}|<|p_c|$. This means that the barrier of $\tau$ is not overpassed at any points because $\dot{u}<v_{\mathrm{van}}$ everywhere in this case. We can conclude that the value smaller than $|p_c|=v_{\mathrm{van}}/v_c^{(-)}$ for $|p_{-\infty}|$ is insufficient to overpass the barrier.

The inhibition of the steady SFP results from the fact that the area with positive $\tau$ increases with time in this case, and the finite stress given at the end-loading point cannot sustain the stable slip. This behavior is qualitatively different from the macroscopic steady slippage, and can be interpreted as creep motion. The transition between the macroscopic steady slippage and the creep motion is understood in a unified way with the present model.

The value $q(X \to \infty)=0$, which gives $\dot{u}(x \to -\infty, t)=v_{\mathrm{van}}$ and $|p_{-\infty}|=|p_c|=v_{\mathrm{van}}/v_c^{(-)}$, is the critical one and it appears at $\mu=\mu_c$. 
The value $|p_c|=v_{\mathrm{van}}/v_c^{(-)}$ gives the lower limit of the end-loading stress required to generate the macroscopic steady slippage. It is to be noted that $v_c^{(-)}$ is the minimum value of $v$. We emphasize that these lower limits can be described in terms of $g$. The case (2) will be treated in numerical simulations in Sec. \ref{secNum2}, and the calculations show the existence of the solutions.

We can also obtain another solution if we put $x_1 =x+vt$, which yields $v=\sqrt{1+a v_{\mathrm{van}}}+\sqrt{a v_{\mathrm{van}}} =\sqrt{1+g}+\sqrt{g} \equiv v_c^{(+)}$. This value is again completely determined by the value of $g$. This solution corresponds to a boundary condition different from that employed here, and we discuss it in Sec. \ref{secLMSH} and Appendix \ref{secAA}.

In the present model, the friction stress is the quadratic form of the slip velocity, Eq. (\ref{eqtau}). We can relax this condition. Based on the barrier picture, it is considered that the steady SFP can exist for the friction law with vanishing friction stress for $\dot{u}$ larger than a critical slip velocity $v_{\mathrm{van}}$. The end-loading stress, which makes the region move with $\dot{u} \ge v_{\mathrm{van}}$ after the slip front has passed, induces the steady SFP and then the macroscopic steady sliding. This condition gives the critical value of the end-loading stress, which determines whether the macroscopic steady slippage occurs or not. The exact form of $\tau(\dot{u})$ quantitatively affects  the critical values of the end-loading stress $p_c$.

The value $v_c^{(\pm)}=\sqrt{1+g} \pm \sqrt{g}$ is shown to be quantitatively valid in Sec. \ref{secLMSH} and Appendix \ref{secAA}, i.e., the SFP velocity is not specific to the friction law with the quadratic form of the slip velocity. The friction law shown in Eq. (\ref{eqtau}) and Fig. \ref{FigCL} is employed because it gives the exact solution for the minimum value of the SFP velocity.

We can extend approximately the result obtained above to the friction law where $\tau$ is finite for $\dot{u}>0$ and $\tau \to 0$ with $\dot{u} \to \infty$. With this friction law, the potential shown in Fig. \ref{FigSq} does not have the constant-value region, and the particle cannot stop. There is no steady state connecting the unstable and stable points in this case. The front cannot propagate steadily and decays, which is the same behavior observed for $q(X \to \infty)>0$ with the friction law (\ref{eqtau}). In terms of the barrier, we can interpret that the width of the barrier is infinitely large in this case, and the slip velocity at the end-loading point never exceeds the barrier. However, if the gradient of the curve in Fig. \ref{FigSq} tends to zero with $q \to -\infty$, we can expect that the slip behavior looks almost steady but with a finite persistent time. This is confirmed by numerical calculations in Sec. \ref{secNum2}.




\subsubsection{Linear Marginal Stability Hypothesis} \label{secLMSH}

We emphasize that the SFP velocities can also be obtained by assuming $\dot{u}(x \to -\infty, t)=v_{\mathrm{van}}$ from the Linear Marginal Stability Hypothesis (LMSH), which has been widely employed to investigate the dynamics of fronts or domain walls propagating spontaneously into an unstable state \cite{Mye1, Sha, Dee, Saa1, Saa2, Lan92}. This leads to the important conclusion that the analytical result obtained in Sec. \ref{secExact2} is not specific to the friction law (\ref{eqtau}), but is applicable to a wide range of friction laws which gives friction stress decreasing to zero linearly at a certain value of the slip velocity. 



Detailed treatment by LMSH is shown in Appendix \ref{secAB}. We here emphasize that we have two solutions for the SFP velocity, $v^{(\pm)}=\sqrt{1+g} \pm \sqrt{g}$, determined by $g=\Big| \partial \tau/\partial \dot{u} |_{\dot{u}=v_{\mathrm{van}}} \Big|$, which is the gradient of $\tau$ at $\dot{u}=v_{\mathrm{van}}$ with vanishing $\tau$. The solution corresponding to the present boundary condition is
\begin{equation}
v^{(-)}=\sqrt{1+g}-\sqrt{g},
\end{equation}
which is the same as the exact analytical solution (\ref{eqv-eg}). The solution $v^{(+)}=\sqrt{1+g}+\sqrt{g}$ corresponds to the front propagation where we arrest the motion at the right end of the block at a certain time with the initial condition $\dot{u}=v_{\mathrm{van}}$ over the whole system. This idea is consistent with the discussion performed in Sec. \ref{secExact2} and Appendix \ref{secAA}.

The important conclusion here is that only the linearized form of the friction law is relevant to the SFP velocity. The only requirement for LMSH is that the friction law can be expanded in powers of $\dot{u}$ around the point where the friction stress vanishes. Based on LMSH, we can conclude that the value of $\partial \tau/\partial \dot{u}$ at this point completely determines the SFP velocity, and the value of the friction stress near $\dot{u}=0$ does not affect the value of the SFP velocity. 

\color{black}

\subsection{Numerical Calculations} \label{secNum2}

\begin{figure}[tbp]
\centering
\includegraphics[width=8.5cm]{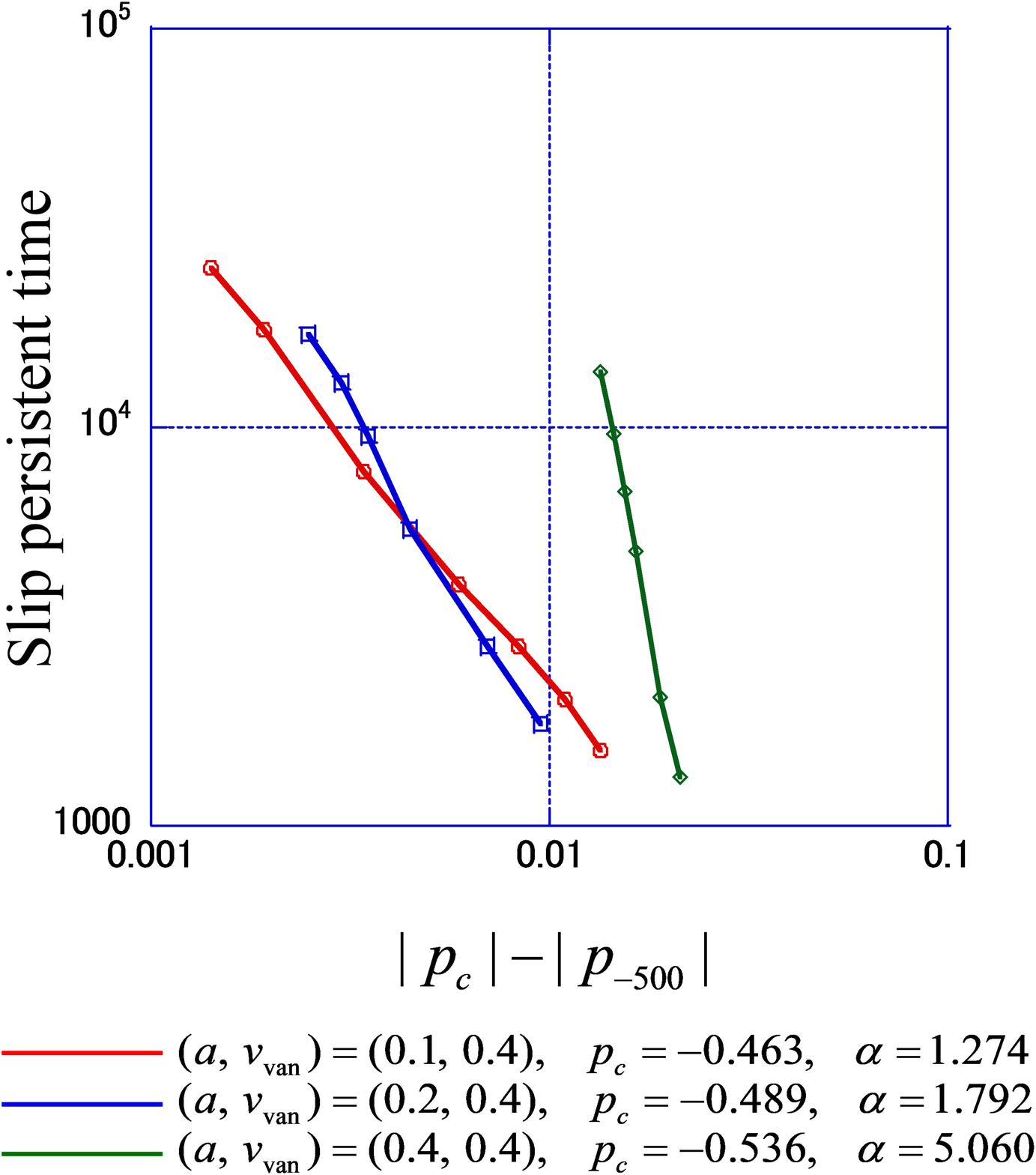}
\caption{(Color online) The slip persistent time in terms of $|p_c|-|p_{-500}|$. The red, blue, and green curves describe the cases $(a, v_{\mathrm{van}}) = (0.1, 0.4), (0.2, 0.4),$ and  $(0.4, 0.4)$, respectively. 
The values of $p_c$ and $\alpha$ are derived by using the least-squares method and approximating the curves by $(|p_c|-|p_{-500})|^{-\alpha}$.}
\label{FigSD}
\end{figure}

In this subsection, we discuss numerical calculations based on the present visco-elastic model. For the calculations, we employ Eqs. (\ref{eqeom1}) and (\ref{eqtau}) at first. This is because the exact solution for the minimum value of the SFP velocity has been obtained, and comparison between the analytical and numerical results is possible. The size of the system in numerical calculations is finite. Here, we take the end-loading point $x=-500$ as the point $x \to -\infty$, so that $p_{-\infty}$ in the analytical discussion is replaced by $p_{-500}$. The right end point is set to be at $x=22500$. Note that within computational limit, the location of the right end point should be so far that the slip is negligible there and the system can be regarded as an infinite medium to justify the analytical result. The results below are confirmed not to depend on its exact location within the computational time. First, we obtain the condition on $p_{-500}$ that is necessary for steady SFP by considering the slip persistent time as a function of $|p_{-500}|$ [Fig. \ref{FigSD}. The figure shows the three cases $(a, v_{\mathrm{van}})=(0.1, 0.4), (0.2, 0.4),$ and $(0.4, 0.4)$]. We define the slip persistent time as the time at which the slip velocity at all points on the slip plane becomes smaller than $0.1 \times |p_{-500}|$ as measured from the onset of end-loading. We have confirmed that the value $0.1$ does not affect the results. If the slip persistent time is finite, steady SFP does not occur. Figure \ref{FigSD} indicates that the slip persistent time diverges as the power law  $(|p_c|-|p_{-500}|)^{-\alpha}$ in the region $|p_{-500}|<|p_c|$, where $\alpha$ is the critical exponent. Note that $p_c$ is obtained numerically here, and it may not be exactly the same as the analytically obtained one. This result indicates that steady SFP exists only in the region $|p_{-500}|>|p_c|$. 
%

\begin{figure}[b]
\centering
\includegraphics[width=8.5cm]{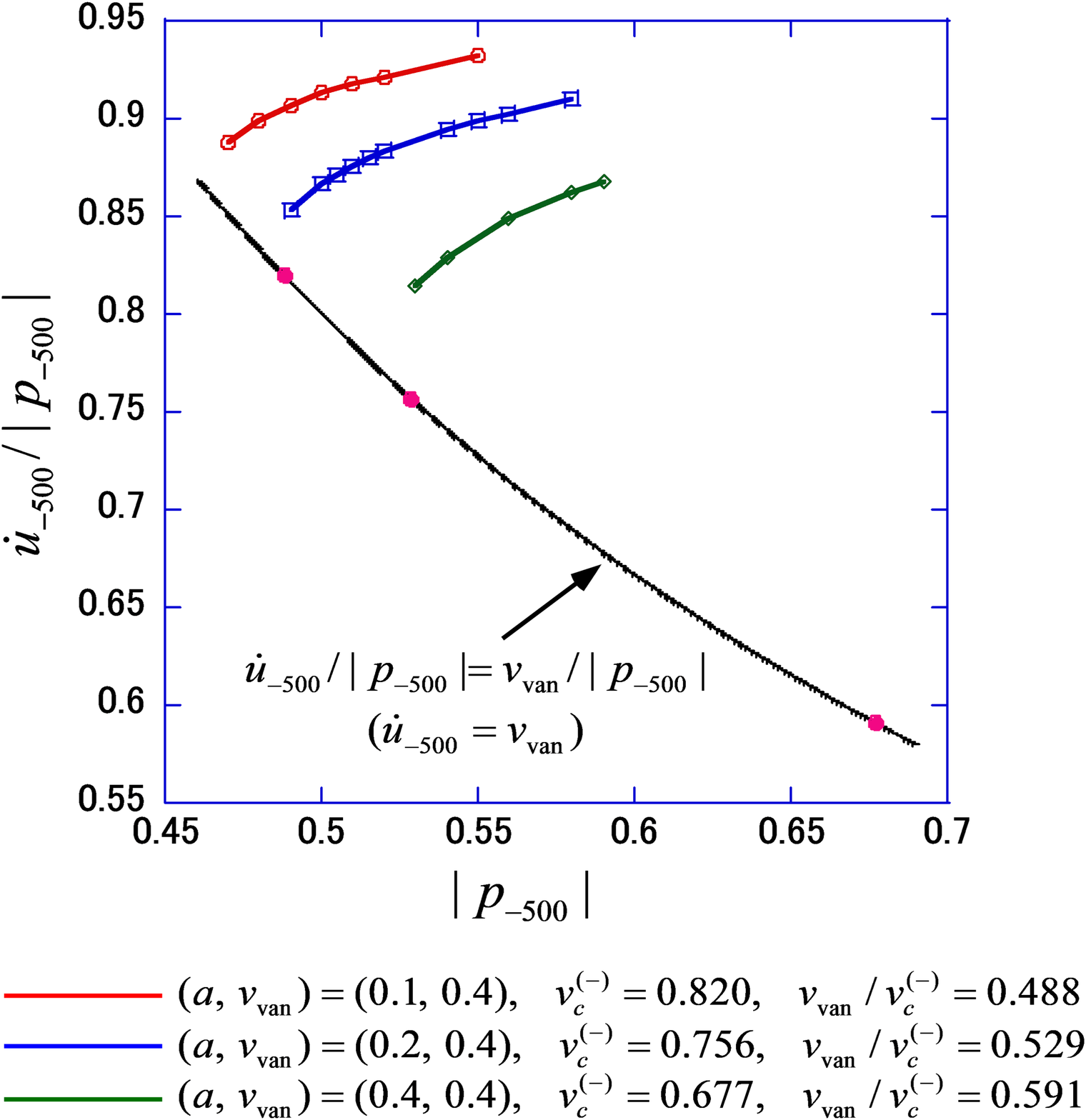}
\caption{(Color online) The value of $\dot{u}_{-500}/|p_{-500}|$. The red, blue and green curves illustrate the cases $(a, v_{\mathrm{van}})=(0.1, 0.4), (0.2, 0.4),$ and $(0.4, 0.4)$, respectively. The thin black solid line indicates the relationship $\dot{u}_{-500}/|p_{-500}|=v_{\mathrm{van}}/|p_{-500}|$. The analytic results give $v_c^{(-)}=\sqrt{1+a v_{\mathrm{van}}}-\sqrt{a v_{\mathrm{van}}}$ and $|p_c|=v_{\mathrm{van}}/v_c^{(-)}=v_{\mathrm{van}}/(\sqrt{1+a v_{\mathrm{van}}}-\sqrt{a v_{\mathrm{van}}})$. These values are given for each set of $(a, v_{\mathrm{van}})$ above and are shown in the figure by the purple points.}
\label{FigVL}
\end{figure}

\begin{figure}[tbp]
\centering
\includegraphics[width=7.5cm]{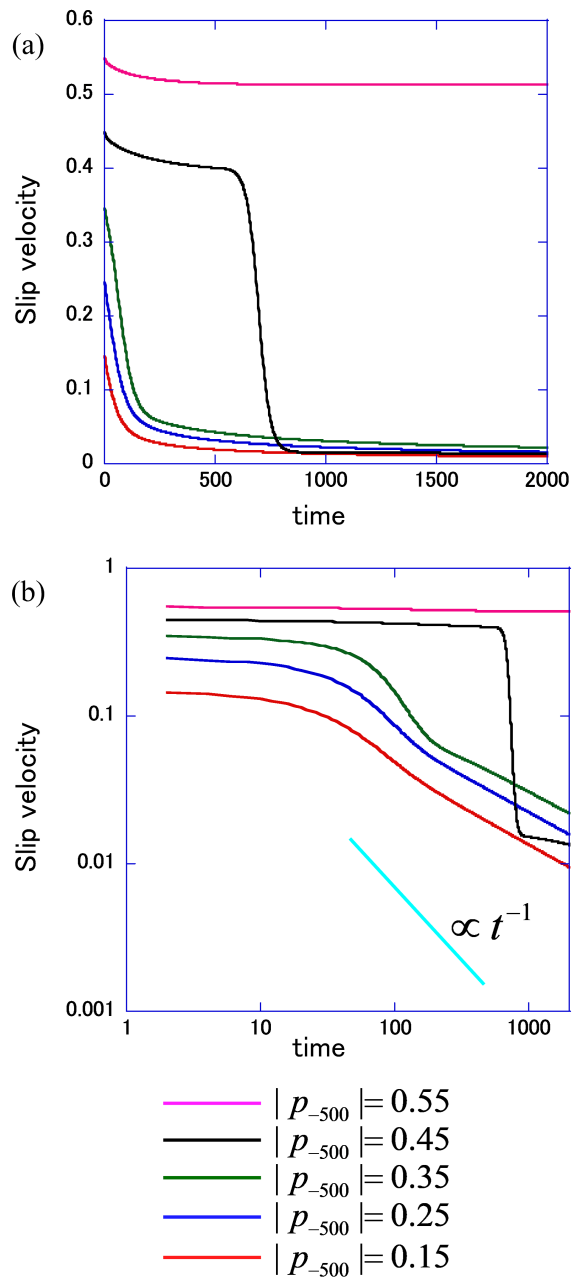}
\caption{(Color online) Temporal changes in the slip velocity at the loading point $x=-500$. 
The parameter set $(a, v_{\mathrm{van}})$ is (0.1, 0.4). The cases $|p_{-500}|=0.15, 0.25, 0.35, 0.45,$ and $0.55$ are calculated. (a) Linear scale, (b) Log-log scale. The short, straight, light-blue line in (b) illustrates a relationship proportional to $t^{-1}$.}
\label{FigDec2}
\end{figure}

\begin{figure}[tbp]
\centering
\includegraphics[width=8.5cm]{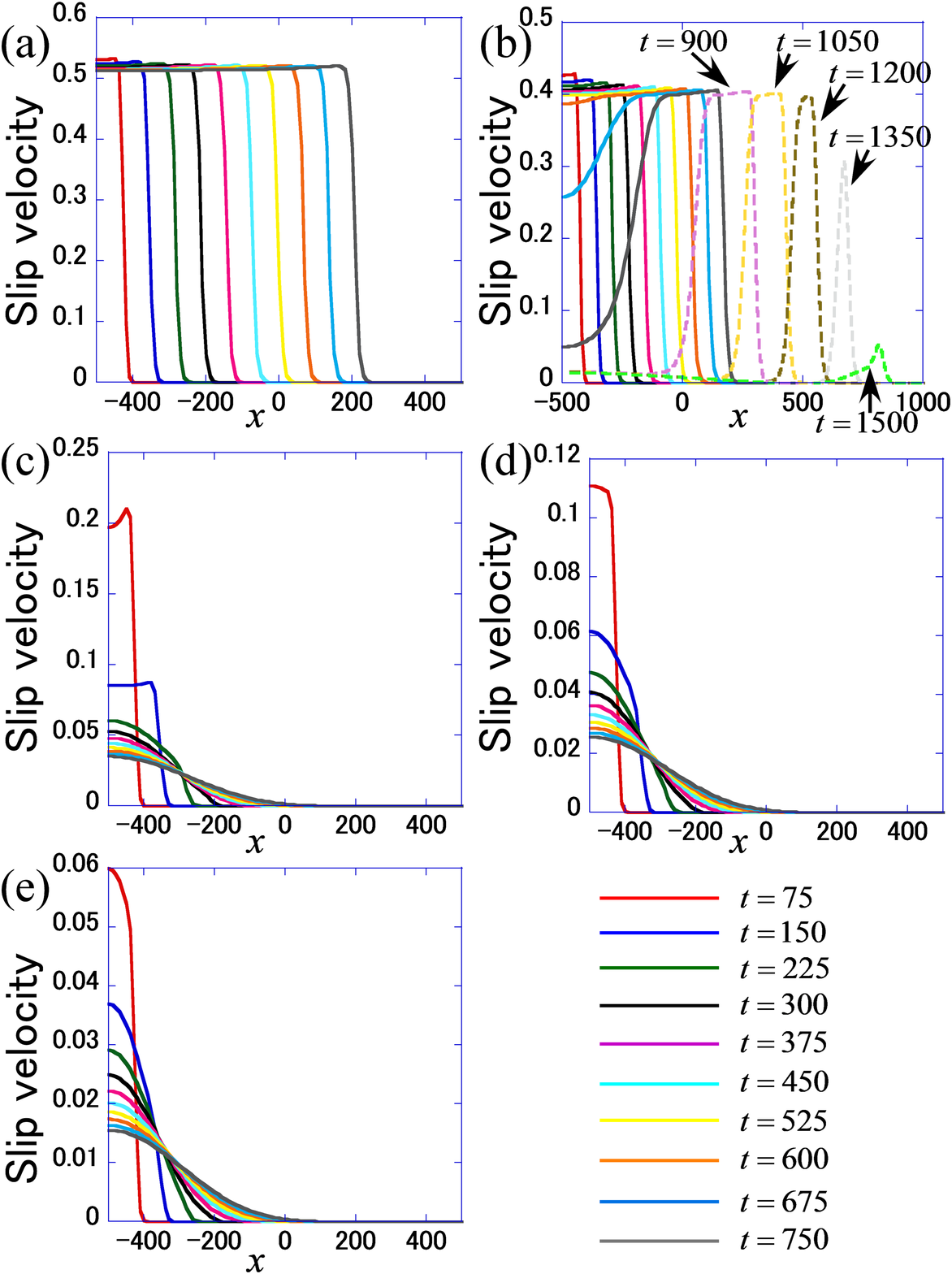}
\caption{(Color online) Spatiotemporal changes in the slip velocity profile.  The parameter set $(a, v_{\mathrm{van}})$ is (0.1, 0.4). 
The five cases $|p_{-500}|=$ (a) $0.55$, (b) $0.45$, (c) $0.35$, (d) $0.25$, and (e) $0.15$ are shown. Lines of different colors show the spatial profiles of the slip velocity at different times. Note that we have performed the calculations up to $t=1500$ only for panel (b).}
\label{FigSTV2}
\end{figure}

We now investigate the SFP velocity in detail in the region $|p_{-500}|>|p_c|$. 
We show the value of $\dot{u}_{-500}/|p_{-500}|$ in the steady state in Fig. \ref{FigVL}, where $\dot{u}_{-500} \equiv \lim_{t \to \infty} \dot{u}(-500,t)$. 
This value describes the SFP velocity in the steady state because the relationship $\dot{u}=\partial u/ \partial t =-v \partial u/ \partial x=-v p$ is satisfied in a steady state moving with the constant velocity $v$. 
The figure shows that the SFP velocity in the present case is less than the SFP velocity in the absence of friction, which is unity here. There exists a minimum value of the SFP velocity. As shown in Sec. \ref{secAna}, the minimum velocity given by the analytical treatment is $v_c^{(-)}=\sqrt{1+a v_{\mathrm{van}}}-\sqrt{a v_{\mathrm{van}}}$. The numerical result for the minimum velocity is consistent with the analytic one. The difference between the numerical and analytic results may come from the finiteness of the system and of the computational time. As Figs. \ref{FigDec2} and \ref{FigSTV2} show, we can also confirm steady SFP for the case $|p_{-500}|=0.55$.

Steady SFP does not exist for $|p_{-500}|<|p_c|$ [Figs. \ref{FigSTV2}(b-e)]. For example, if $|p_{-500}|=0.45 < |p_c| =0.461$, as for the case shown in Fig. $\ref{FigSTV2}$ (b), the slip velocity decreases and approaches zero with increasing time, and non-steady pulse-like slip behavior appears. For cases with $|p_{-500}|<|p_c|$, $\dot{u}$ and $p$ approach zero and $p_{-500}$, respectively, in the whole space.

We can understand whether steady SFP emerges in terms of  $p_{-500}$ based on these results. For the case $|p_{-500}| > |p_c|$, steady SFP occurs, and this slip propagates into the entire system with a finite velocity. On the other hand, if $|p_{-500}| < |p_c|$, $\dot{u}$ approaches zero over the whole plane with increasing time, indicating that there is no macroscopic steady slippage. However, the slip diverges with $t \to \infty$ in this case, which is clear from Fig. \ref{FigDec2}(b). The slip velocity decays more slowly than $t^{-1}$, and the cumulative slip does not converge. This behavior can be interpreted as the creep motion. These statements indicate that creep motion and macroscopic steady slippage can be modeled in the single framework, and they are categorized in terms of $|p_{-500}|$. The magnitude of the critical end-loading stress at $x \to -\infty$, $|p_c|$, is the lower limit of the end-loading stress that generates steady SFP, and the value is almost quantitatively equal to the analytic one.

\begin{figure}[tbp]
\centering
\includegraphics[width=8.5cm]{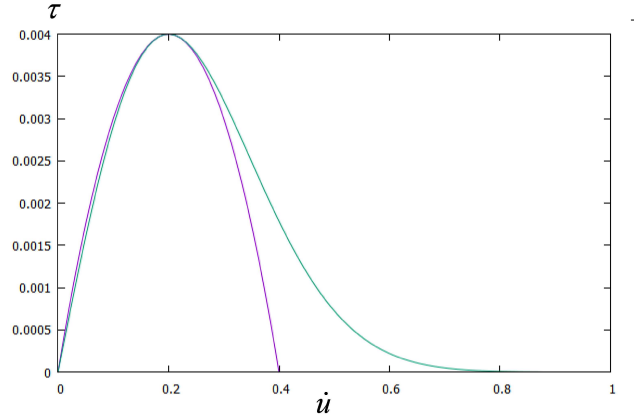}
\caption{(Color online) Constitutive law of the local friction stress. The law (\ref{eqCLe}) with $(a, 2b)=(0.1, 0.4)$ is shown by the green curve, while the law (\ref{eqtau}) with $(a, v_{\mathrm{van}})=(a, 2b)=(0.1, 0.4)$ is drawn by the purple curve.}
\label{FigCL2}
\end{figure}

\begin{figure}[tbp]
\centering
\includegraphics[width=8.5cm]{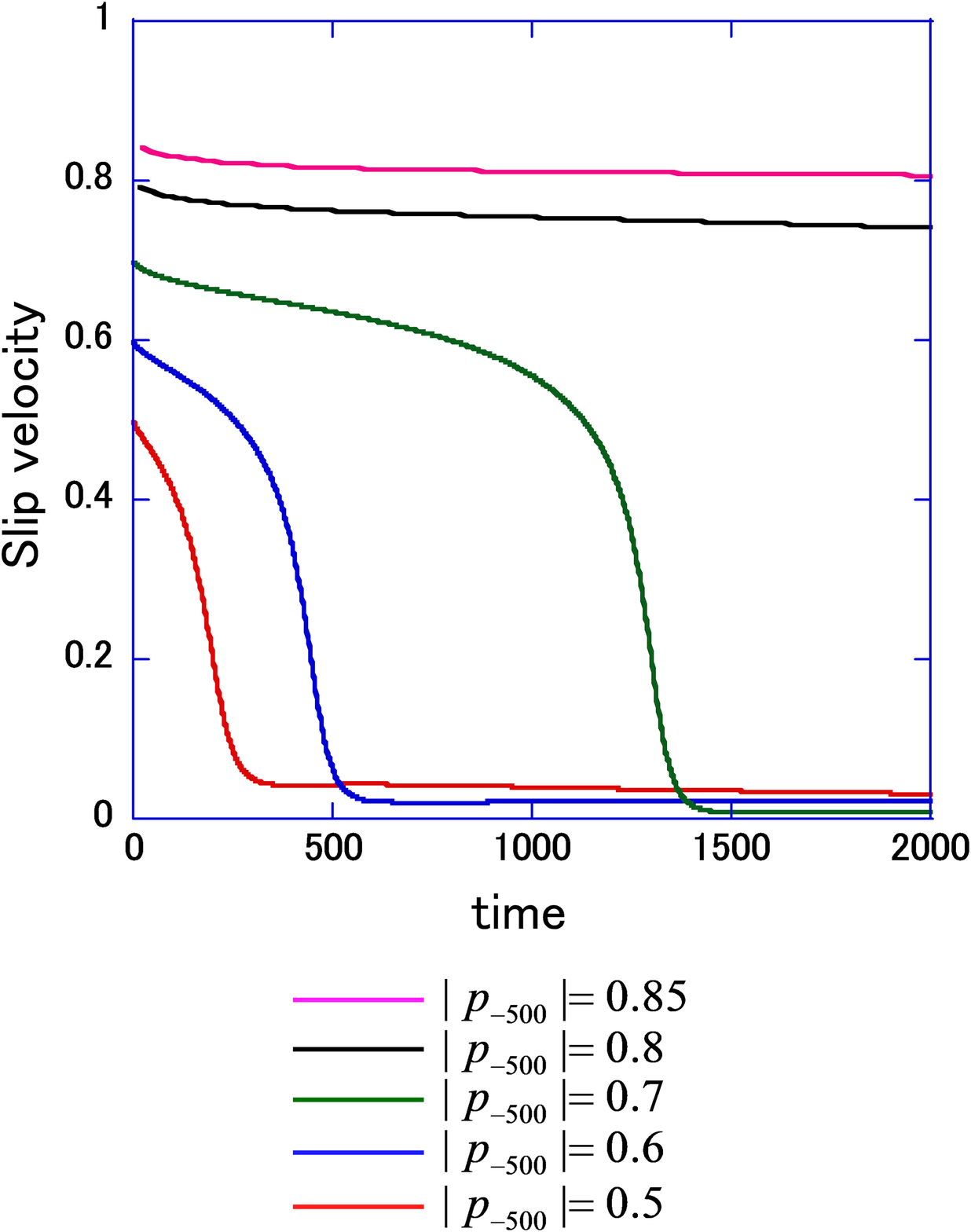}
\caption{(Color online) Temporal changes in the slip velocity at the loading point $x=-500$ with the friction law (\ref{eqCLe}). The parameter set $(a, b)$ is (0.1, 0.2). The cases $|p_{-500}|=0.5, 0.6, 0.7, 0.8$ and $0.85$ are calculated.}
\label{FigDec3}
\end{figure}

\begin{figure}[tbp]
\centering
\includegraphics[width=8.5cm]{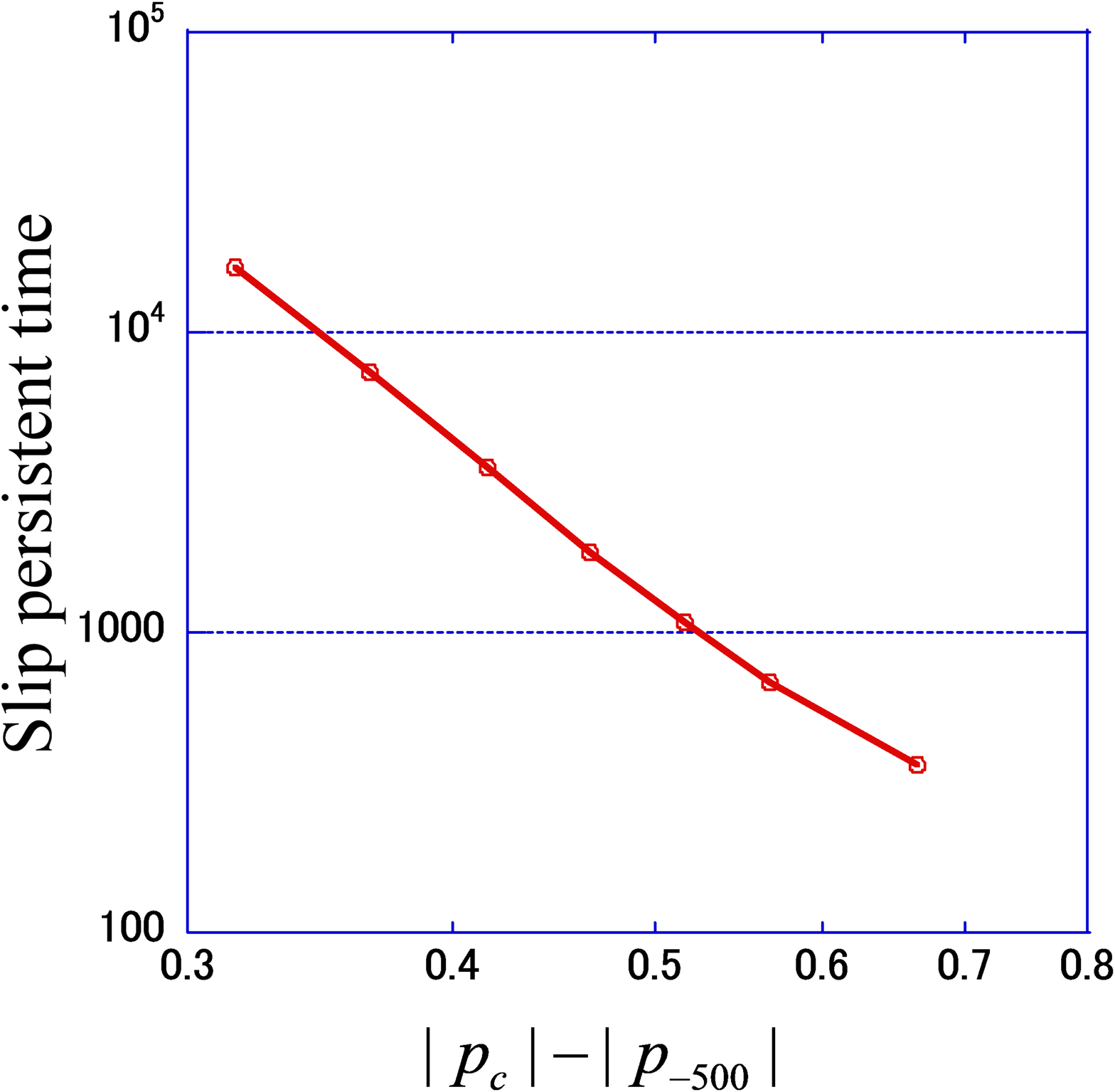}
\caption{(Color online) The slip persistent time in terms of $|p_c|-|p_{-500}|$. The parameters $(a, b) = (0.1, 0.2)$ are assumed. The values of $p_c$ and $\alpha$ are $-1.166$ and $5.559$, respectively.}
\label{FigSD2}
\end{figure}

We finally confirm that the discussion for the friction law (\ref{eqtau}) can be extended to the case where  $\tau$ is always positive for $\dot{u}>0$ and $\tau \to 0$ with $\dot{u} \to \infty$. We assume the friction law shown in Fig. \ref{FigCL2}, which has the form
\begin{equation}
\tau =ab e^{1/2} \dot{u} e^{-\dot{u}^2/2b^2}. \label{eqCLe}
\end{equation}
The temporal change in the slip velocity at the end-loading point with this friction law is illustrated in Fig. \ref{FigDec3}, which shows qualitatively similar behavior to Fig. \ref{FigDec2}(a). To investigate dependence of the slip behavior on $|p_{-500}|$, we show the slip persistent time in Fig. \ref{FigSD2}. Actually, the slip persistent time never diverges with the friction law (\ref{eqCLe}), since the steady state does not emerge as shown in Sec. \ref{secExact2}. However, Fig. \ref{FigSD2} shows the power law within the time range investigated, and distinction between Figs. \ref{FigSD} and \ref{FigSD2}  is impossible within the computational limit. These results clearly show that apparent critical behavior emerges with the friction law (\ref{eqCLe}). The numerical results based on the laws (\ref{eqtau}) and (\ref{eqCLe}) are quantitatively different. In particular, $|p_c|$ for the law (\ref{eqCLe}) is larger than that for (\ref{eqtau}). This is due to the fact that the barrier described by (\ref{eqCLe}) is larger than that by (\ref{eqtau}), as shown in Fig. \ref{FigCL2}. We can conclude that implications based on the friction law (\ref{eqtau}) is qualitatively applicable to the law where the friction stress decreases with increasing slip velocity in the large slip velocity regime, but never vanishes.

We finally apply the analytical result obtained for the friction law (\ref{eqtau}) to the numerical result for Eq. (\ref{eqCLe}). The relation $v_c^{(-)}=v_{\mathrm{van}}/|p_c|$ was obtained for Eq. (\ref{eqtau}). The value $v_{\mathrm{van}}$ is the width of the barrier illustrated by the purple curve in Fig. \ref{FigCL2}. The width of the green barrier in Fig. \ref{FigCL2} is estimated as 0.8. The value $|p_c|$ for Eq. (\ref{eqCLe}) was estimated at 1.166 [see the caption for Fig. \ref{FigSD2}]. We then have $0.8/1.166=0.686$. Actually, the SFP velocity in the case shown by the left end point in Fig. \ref{FigSD2} was roughly evaluated as 0.762, which is almost consistent with 0.686.

\color{black}

\section{Discussion and Conclusions} \label{secDisCon}

We examined a 1D visco-elastic model under end-loading stress to investigate the macroscopic slip initiation. We assumed the local friction law that has a quadratic form of the slip velocity and gives vanishing friction stress above a certain velocity. We analytically obtained two kinds of the critical slip-front-propagation velocities. One of the velocities corresponds to the boundary condition used here, and it gives the minimum value for the SFP velocity. We also obtained the lower limit of the end-loading stress required at the loading point to cause the macroscopic steady slippage of the block. If the end-loading stress is larger (smaller) than the limit, the macroscopic steady slippage (creep motion) emerges.
Note that the analytical solution for the SFP velocity associated with the boundary condition used here and the lower limit of the end-loading stress are consistent with those obtained numerically. 


We extended the framework to more general friction laws. First, the friction law was extended to laws with non-quadratic dependence on the slip velocity. The only requirement to the friction law is that the friction stress vanishes with $\dot{u}$ larger than a certain critical value. The Linear Marginal Stability Hypothesis gives the solutions for the SFP velocities same as the exact ones. We emphasize that the value $\partial \tau/\partial \dot{u}$ at the critical value completely determines the SFP velocity. Second, we also showed that the analytical result obtained for the quadratic-form law is approximately valid even in the friction law where the friction stress approaches zero with $\dot{u} \to \infty$. We gave analytical conclusion for this extension, and numerical simulations are found to support such suggestions. These statements lead to the conclusion that the detail of the friction law does not affect the SFP velocity.


We have also obtained some implications for the creep motion. Such creep motion may be related to precursors observed prior to macroscopic steady slippage, because the creep motion can be arrested easily by negligibly small perturbations of the stress on the plane, and it leaves residual stress there after end-loading stress is removed. Such residual stress may induce macroscopic steady slippage when the end-loading stress smaller than the critical value is applied. We plan to investigate this possibility in future work. 

We can also draw some seismological implications from the results obtained here. For example, the pulse-like slippage shown in Fig. \ref{FigSTV2}(b) may explain the slip behavior observed for ordinary earthquakes \cite{Wal}. Although pulse-like slippage has been explained in terms of, e.g., the dilatancy effect of the fault rocks \cite{Suz08}, our results show that the nonlinear friction law itself can generate such slippage. Furthermore, note that if $|p_{-\infty}| \ll |p_c|$, the emergent slip velocity is negligibly small. Slow earthquakes shows this behavior. Thus, we have succeeded in dynamically modeling slow earthquakes, at least qualitatively, in terms of a friction law that depends nonlinearly on the slip velocity. 

\acknowledgments
This research was supported by JSPS KAKENHI Grant Number JP26400403 and JP16K17795 in Scientific Research on Innovative Areas ``Science of Slow Earthquakes''.
T. S. is also supported by JSPS KAKENHI Grant Number JP16H06478.
H. M. is also supported by JSPS KAKENHI Grant Number JP17K05586.
This study was supported by ERI JURP 2018-G-04.
This work was partially supported by Aoyama Gakuin University-Supported Program ``Promotion for Ongoing Research Program''.


\appendix

\renewcommand{\theequation}{A\arabic{equation}}
\setcounter{equation}{0}

\section{The Physical Meaning of Another Solution for the SPF Velocity} \label{secAA}

As mentioned in the text, putting $x_1=x+vt$ generates the solution $v=\sqrt{1 + a v_{\mathrm{van}}} +\sqrt{a v_{\mathrm{van}}} \equiv v_c^{(+)}$. Here, we discuss the physical interpretation of this solution. With $x_1=x+vt$, we obtain $p(x \to -\infty, t)=v_{\mathrm{van}}/v_c^{(+)}$ and $p(x \to \infty, t)=0$. 
This steady state is realized by fixing the stress to be $v_{\mathrm{van}}/v_c^{(+)}$ over the whole system in the initial state and letting $p(x \to \infty, t)=0$ after $t=t_0$. 
The slip front propagates to the left with the constant speed $v_c^{(+)}$. In terms of the slip velocity, we have $\dot{u}=v_{\mathrm{van}}$ over the whole system in the initial state. If we make the slip velocity zero at the right end of the block at $t=t_0$, we obtain a steady state after $t \to \infty$. Though this case does not coincide with the situation here, the velocity $v_c^{(+)}$ actually has  physical meaning.

\renewcommand{\theequation}{B\arabic{equation}}
\setcounter{equation}{0}

\section{Introduction of the Linear Marginal Stability Hypothesis} \label{secAB}

Here, we explain the procedure for applying the LMSH. First, we define $s$ as a nondimensional variable characterizing the state of the system$-$for example, the normalized slip distance or the slip velocity for the motion of the continuum$-$and we consider the dynamics of the spontaneous propagation of $s$. We consider a 1D system, and treat the front intruding into the unstable region. We consider two cases: one where $s=0$ is stable and this region intrudes into the unstable region with $s \neq 0$, and the other where $s=0$ is unstable and the stable region with $s \neq 0$ intrudes into this unstable region. Here, the front for the former solution is called an ``extruding front,'' and that for the latter one is called an ``intruding front.'' The front is mathematically defined to be located where only terms $O(|s|)$ play important roles and terms $O(|s|^2)$ become negligible in the governing equations. This definition is conistent with that in Sec. \ref{secExact2}, as we show later in this Appendix. 

\begin{figure}[tbp]
\centering
\includegraphics[width=8.5cm]{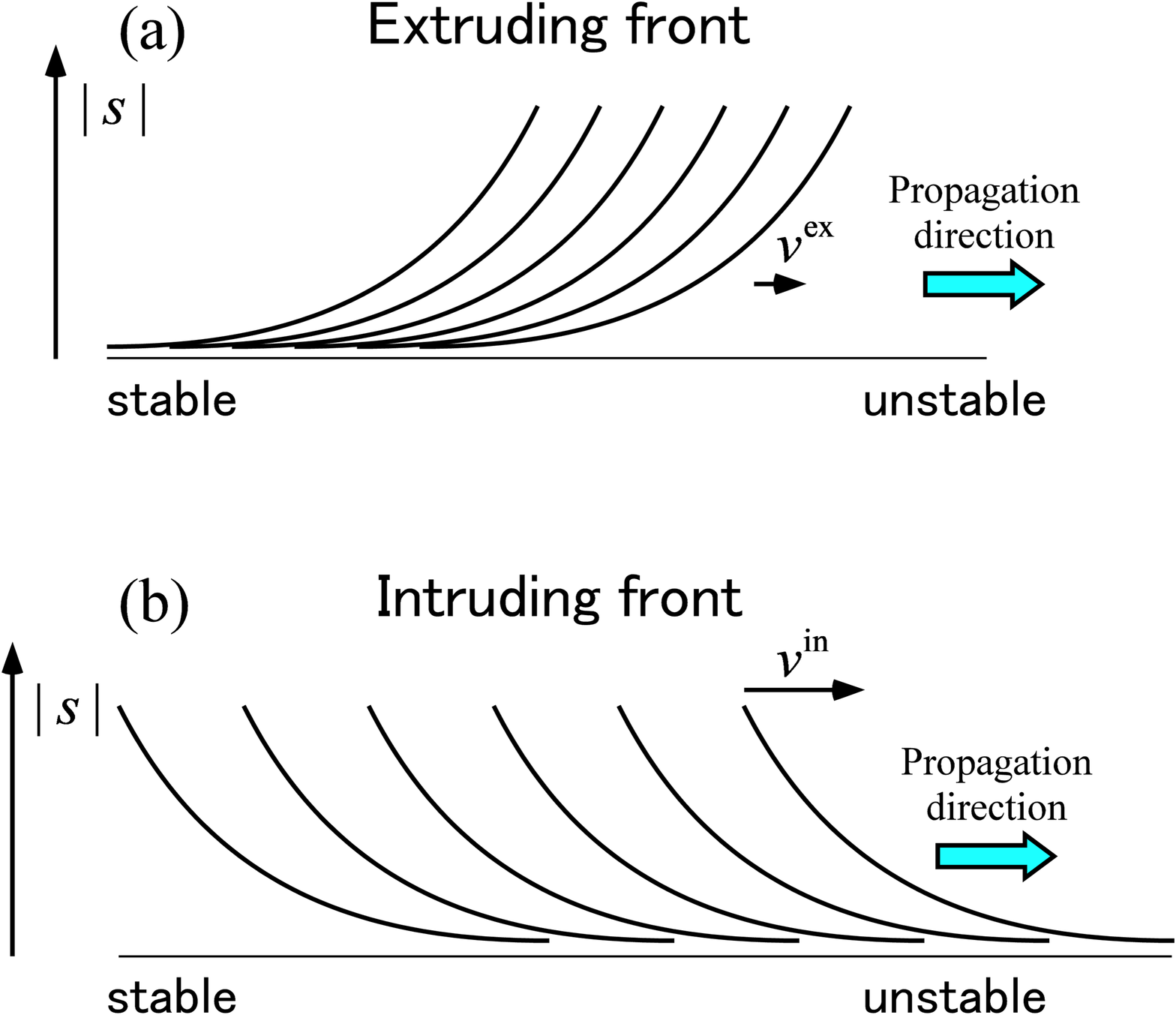}
\caption{(Color online) Schematic representations of (a) an extruding front and (b) an intruding front in terms of $|s|$.}
\label{FigSch}
\end{figure}

We assume the front of $s$ to be the plane wave $s \sim \exp[\mp i(kx-\omega t)]$, where the frequency $\omega$ and wave number $k$ are both complex. This description yields $|s|=\exp [\pm (k_i x - \omega_i t)]$, where $k_i$ and $\omega_i$ are the imaginary parts of $k$ and $\omega$, respectively, and are assumed to be nonnegative. Note that the fronts $\exp(k_i x -\omega_i t)$ and $\exp[-(k_i x -\omega_i t)]$ describe the extruding and intruding fronts, respectively, based on their definition (Fig. \ref{FigSch}). We have four unknown parameters: $k_i$, $\omega_i$, $k_r$ (the real part of $k$), and $\omega_r$ (the real part of $\omega$). The parameters $k_r$ and $\omega_r$ are also assumed to be nonnegative. These four parameters can be determined from the viewpoint of LMSH by four independent equations: the real and imaginary parts of the dispersion relation, the growth stability and the propagation stability. 

Here, we provide detailed explanations for the growth and propagation stabilities. First, growth stability corresponds to the condition
\begin{equation}
\frac{\partial \omega_i}{\partial k_r}=0.  \label{eqABGI1}
\end{equation}
To understand the physical meaning of this relationship, let us assume that the wave number has small disturbance, $k=k_0 + \Delta k \ ( \Delta k \in \mathbb{R})$, where $k_0$ is a constant complex number. Let us take into account of this assumption and the following relationship:
\begin{eqnarray}
|\exp (\pm i(kx-\omega t))| &=& |\exp(\pm i ((k_r + i k_i)x - (\omega_r + i \omega_i)t))| \nonumber \\
&=& |\exp(\mp (k_i x - \omega_i t)) \exp(\pm i (k_r x - \omega_r t))| \nonumber \\
&\sim& \exp(\mp(k_i x - \omega_{i0} t) \exp(\pm \frac{\partial \omega_i}{\partial k_r} \Delta k \cdot t), \label{eqABAm}
\end{eqnarray}
where $k_{i0}$ and $\omega_{i0}$ are the imaginary parts of $k_0$ and $\omega(k=k_0)$, respectively. If $\partial \omega_i / \partial k_r \ne 0$, we can select $\Delta k$ satisfying $(\partial \omega_i / \partial k_r) \Delta k > 0$, which induces an exponential increase in the amplitude of the disturbance. Such an increase does not generate steady front propagation, so that Eq. (\ref{eqABGI1}) assures stability of the growth of disturbance.

Second, the propagation stability gives the relationship
\begin{equation}
\frac{\omega_i}{k_i} = \frac{\partial \omega_i}{\partial k_i} = c, \label{eqABPI1}
\end{equation}
where $c$ is the front propagation velocity. Equation (\ref{eqABPI1}) clearly shows that the phase velocity $c_p =\omega_i/k_i$ equals the group velocity $c_g =\partial \omega_i/\partial k_i$. Since the disturbance propagates with the group velocity, this condition shows that the disturbance and the front propagate with the same velocity. However, it is important to note that for stability, the relationship $c_p \ge c_g$ is sufficient because the disturbance is overtaken by the front with such a relationship. Nonetheless, Ref. \cite{Saa1} mathematically showed that the relationship (\ref{eqABPI1}) is satisfied for spontaneous front propagation. 

We need two more requirements to determine all the parameters. We assume $\omega$ to be a regular function of $k$, then we have
\begin{equation}
\frac{\partial \omega_r}{\partial k_i} =0, \label{eqABGI2}
\end{equation}
and
\begin{equation}
\frac{\partial \omega_r}{\partial k_r} = c, \label{eqABPI2-1}
\end{equation}
from the Cauchy-Riemann relationship and the growth and propagation stability conditions. Additionally, steady front propagation requires $\partial \omega_r/ \partial k_r = \omega_r / k_r$, because the front and the disturbance must propagate with the same velocity also for the real part of $\exp[\pm (kx-\omega t)]$, so that we have
\begin{equation}
\frac{\partial \omega_r}{\partial k_r} =\frac{\omega_r}{k_r} = c. \label{eqABPI2}
\end{equation}

Next, we apply the LMSH to the frictional phenomenon between the substrate and the visco-elastic block assuming the friction law (\ref{eqtau}). 
We expand the friction law appearing in the equation of motion (\ref{eqeom2}) near the point where the friction stress vanishes, i.e., $\dot{u}=v_{\mathrm{van}}$, and assume $\dot{u}-v_{\mathrm{van}}<0$. This is because the region where $\dot{u} \sim v_{\mathrm{van}}$ and $\dot{u}<v_{\mathrm{van}}$ is unstable due to the velocity-weakening behavior (note that the gradient of the curve shown in Fig. \ref{FigCL} near $\dot{u}=v_{\mathrm{van}}$ and $\dot{u}<v_{\mathrm{van}}$ is negative); therefore, the unstable region is overtaken by the region $\dot{u}=v_{\mathrm{van}}$. The slip front exists there, and we have the linear equation
\begin{equation}
\ddot{u}= u'' + \dot{u}'' + a v_{\mathrm{van}} (\dot{u} - v_{\mathrm{van}}), \label{eqN1}
\end{equation}
because $\partial \tau/\partial \dot{u}|_{\dot{u}-v_{\mathrm{van}}=0}=-a v_{\mathrm{van}}$. This linearization is equivalent to neglecting $O(|\dot{u} - v_{\mathrm{van}}|^2)$, consistent with the definition of the front in this subsection. Using $w=u-v_{\mathrm{van}}t$, Eq. (\ref{eqN1}) reduces to
\begin{equation}
\ddot{w}= w'' + \dot{w}'' + a v_{\mathrm{van}} \dot{w}. \label{eqeom3}
\end{equation}
Here we put $s \equiv \dot{w}=\dot{u}-v_{\mathrm{van}}$. Hence the linearization employed in Eq. (\ref{eqN1}) is equivalent to the linearization around $s=0$ noted above. Thus, we obtain
\begin{equation}
\ddot{s}= s'' + \dot{s}'' + a v_{\mathrm{van}} \dot{s}. \label{eqeom4}
\end{equation}
This equation is a basis for the analytical treatment below. Note that this includes only $\partial \tau/\partial \dot{u}|_{\dot{u}=v_{\mathrm{van}}}$ as the value characterizing the system. Therefore, we rewrite Eq. (\ref{eqeom4}) by using $g \equiv \Big|\partial \tau/\partial \dot{u}|_{\dot{u}=v_{\mathrm{van}}} \Big|$ as
\begin{equation}
\ddot{s}= s'' + \dot{s}'' + g \dot{s}. \label{eqeom5}
\end{equation}

\begin{figure}[tbp]
\centering
\includegraphics[width=8.5cm]{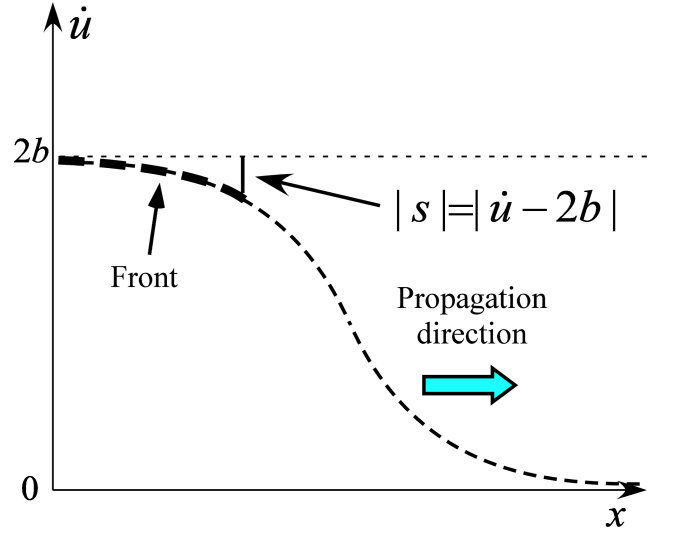}
\caption{(Color online) Schematic representation of the SFP in terms of the slip velocity. Compare the form of the slip front with Fig. \ref{FigSch}(a).}
\label{FigFr}
\end{figure}

Let us begin our analytical treatment with Eq. (\ref{eqeom5}). If we substitute $\exp[-i (k x - \omega t) ]$ into Eq. (\ref{eqeom5}), the dispersion relation is easily shown to be
\begin{equation}
-\omega^2 = - k^2 - i \omega k^2 + ig \omega.
\end{equation}
The real and imaginary parts of the dispersion relation are
\begin{equation}
(\omega_i - 1)(k_r^2 - k_i^2) + 2 \omega_r k_r k_i + (\omega_r^2 - \omega_i^2) - g \omega_i = 0, \label{eqD1}
\end{equation}
\begin{equation}
2 k_r k_i (\omega_i - 1)  - \omega_r (k_r^2 - k_i^2) + 2 \omega_r \omega_i + g \omega_r = 0, \label{eqD2}
\end{equation}
respectively. Differentiating Eqs. (\ref{eqD1}) and (\ref{eqD2}) with respect to $k_r$ and employing the growth and propagation stability conditions [Eqs. (\ref{eqABGI1}) and (\ref{eqABPI2})] yield
\begin{equation}
2 (\omega_i - 1) k_r + 2 k_i (c k_r + \omega_r ) + 2 \omega_r c = 0, \label{eqD1d1}
\end{equation}
\begin{equation}
2 (\omega_i - 1) k_i - c (k_r^2-k_i^2) - 2 \omega_r k_r + 2 \omega_i c + gc =0, \label{eqD2d1}
\end{equation}
respectively. Moreover, differentiating Eqs. (\ref{eqD1}) and (\ref{eqD2}) with respect to $k_i$ and employing the propagation and growth stability conditions [Eqs. (\ref{eqABPI1}) and (\ref{eqABGI2})] give
\begin{equation}
c (k_r^2-k_i^2) -2(\omega_i - 1) k_i + 2 \omega_r k_r -2 \omega_i c - gc =0, \label{eqD1d2}
\end{equation}
\begin{equation}
2 \omega_i k_r + 2(\omega_i - 1) k_r + 2 \omega_r k_i + 2 \omega_r c =0, \label{eqD2d2}
\end{equation}
respectively. We also used the relationship $\partial \omega_i/\partial k_i =c$. We have four unknown variants $k_r, k_i, \omega_r, \omega_i$, though there exist six equations (\ref{eqD1})-(\ref{eqD2d2}). However, because Eqs. (\ref{eqD1d1}) and (\ref{eqD2d2}), and Eqs.  (\ref{eqD2d1}) and (\ref{eqD1d2}), respectively, are exactly the same, the independent equations are (\ref{eqD1}), (\ref{eqD2}), (\ref{eqD1d1}) and (\ref{eqD2d1}). In addition, Eq. (\ref{eqD2}) and Eq. (\ref{eqD2d1}) give $2 \omega_r k_r^2=0$, which yields $k_r=0$ or $\omega_r=0$. Moreover, if $k_r=0$ or $\omega_r=0$, we can conclude that $k_r=\omega_r=0$ based on the propagation stability (\ref{eqABPI2}).

Employing $k_r = \omega_r =0$, the left hand sides of Eqs. (\ref{eqD2}) and (\ref{eqD1d1}) are identically zero. We also have
\begin{equation}
(- \omega_i + 1) k_i^2 - \omega_i^2 - g \omega_i=0, \label{eqI1}
\end{equation}
\begin{equation}
2 (\omega_i - 1) k_i + c k_i^2 +2 \omega_i c + gc =0, \label{eqI2}
\end{equation}
from Eqs. (\ref{eqD1}) and (\ref{eqD2d1}), respectively. In addition, dividing Eq. (\ref{eqI1}) by $k_i$ and employing the relationship $\omega_i/k_i = c$, we obtain the equation
\begin{equation}
(- \omega_i +1) k_i - \omega_i c - gc = 0. \label{eqI3}
\end{equation}
We obtain $k_i, \ \omega_i, \ c$ from Eqs. (\ref{eqI1}), (\ref{eqI2}), and (\ref{eqI3}). 
First, Eqs. (\ref{eqI2}) and (\ref{eqI3}) give
\begin{equation}
k_i^2 = g. \label{eqks1}
\end{equation}
We have $k_i = \sqrt{g}$ from this equation since $k_i$ is assumed to be positive. With this result and Eq. (\ref{eqI1}), we can see that $\omega_i$ obeys the equation
\begin{equation}
\omega_i^2 + 2g \omega_i -g = 0, \label{eqo2}
\end{equation}
which gives the solution
\begin{equation}
\omega_i = - g \pm \sqrt{g^2 + g }. \label{eqOmega}
\end{equation}
We select the plus sign and write the solution
\begin{equation}
\omega_i = \sqrt{g^2 + g} - g,
\end{equation}
since $\omega_i$ is assumed to be positive. This equation, together with Eq. (\ref{eqks1}), gives the SFP velocity $c$ in the form
\begin{eqnarray}
c &=& \frac{\omega_i}{k_i} = \left( \sqrt{ g^2 + g} - g \right) \sqrt{\frac{1}{g}} \nonumber \\
&=& \sqrt{1 + g} - \sqrt{g}. \label{eqc}
\end{eqnarray}
Thus, we have $v^{\mathrm{ex}}=\sqrt{1+g}-\sqrt{g}$, which is smaller than the elastic-wave velocity and exactly the same as $v_c^{(-)}$.

The slip front treated in Sec. \ref{secExact2} was an extruding front, which is consistent with the statement that $v^{\mathrm{ex}}=v_c^{(-)}$. See Fig. \ref{FigFr}; note that the friction stress reaches a critical value after reaching its peak value $ab^2$ where $|s|$ is much smaller than $v_{\mathrm{van}}$. On the other hand, as mentioned in Sec. \ref{secLMSH}, $v_c^{(+)}$ describes the SFP velocity for the case in which the slip velocity is initially $v_{\mathrm{van}}$ over the whole region and we arrest the slip at $x \to \infty$. Note here that if the propagation direction is leftward in Fig. \ref{FigSch}(a), we can confirm that a slip front will emerge with velocity $v^{\mathrm{in}}$ [see also Fig. \ref{FigSch}(b)]. Additionally, if we substitute $\exp [i (k x -\omega t)]$ into Eq. (\ref{eqeom5}), we easily obtain $v^{\mathrm{in}}=\sqrt{1+g}+\sqrt{g}$, which yields $v^{\mathrm{in}}=v_c^{(+)}$.




\end{document}